\documentclass[conference]{IEEEtran}

\usepackage{cite}
\usepackage{amsmath,amssymb,amsfonts}
\usepackage{algorithmic}
\usepackage{graphicx}
\usepackage{textcomp}
\usepackage{xcolor}
\usepackage{comment}
\def\BibTeX{{\rm B\kern-.05em{\sc i\kern-.025em b}\kern-.08em
    T\kern-.1667em\lower.7ex\hbox{E}\kern-.125emX}}

\begin{document}

\title{Automated Code Review Using Large Language Models at Ericsson: An Experience Report}


\author{\IEEEauthorblockN{Shweta Ramesh}
\IEEEauthorblockA{
\textit{Ericsson}\\
Bangalore, India  \\
Shweta.Ramesh@ericsson.com}
\and
\IEEEauthorblockN{Joy Bose}
\IEEEauthorblockA{
\textit{Ericsson}\\
Bangalore, India  \\
joy.bose@ericsson.com}
\and
\IEEEauthorblockN{ Hamender Singh}
\IEEEauthorblockA{
\textit{Ericsson}\\
Bangalore, India  \\
hamender.singh@ericsson.com}
\and
\IEEEauthorblockN{ Ak Raghavan}
\IEEEauthorblockA{
\textit{Ericsson}\\
Bangalore, India \\
ak.raghavan@ericsson.com}
\and
\IEEEauthorblockN{Sujoy Roy Chowdhury}
\IEEEauthorblockA{
\textit{Ericsson}\\
Bangalore, India \\
sujoy.chowdhury@ericsson.com}
\and
\IEEEauthorblockN{Giriprasad Sridhara}
\IEEEauthorblockA{
\textit{Ericsson}\\
Bangalore, India \\
giriprasad.sridhara@ericsson.com}
\and
\IEEEauthorblockN{Nishrith Saini}
\IEEEauthorblockA{
\textit{Ericsson}\\
Stockholm, Sweden \\
nishrith.saini@ericsson.com
}
\and
\IEEEauthorblockN{Ricardo Britto}
\IEEEauthorblockA{
\textit{Ericsson}\\
Stockholm, Sweden \\
ricardo.britto@ericsson.com}

}

\maketitle

\begin{abstract} 
Code review is one of the primary means of assuring the quality of released software along with testing and static analysis. However, code review requires experienced developers who may not always have the time to perform an in-depth review of code. Thus, automating code review can help alleviate the cognitive burden on experienced software developers allowing them to focus on their primary activities of writing code to add new features and fix bugs. In this paper, we describe our experience in using Large Language Models towards automating the code review process in Ericsson. We describe the development of a lightweight tool using LLMs and static program analysis. We then describe our preliminary experiments with experienced developers in evaluating our code review tool and the encouraging results.
\end{abstract}


\section{Introduction}
There are multiple orthogonal and mutually complementary techniques to ensure the quality of the software. Typically, these methods include software testing that dynamically executes the software under test to detect bugs; static program analysis that analyzes code without running it to discover problems such as a potential division by zero; and code reviews which involve experienced developers reading the code to find issues in the software.

Code reviews are critical to ensuring the quality and maintainability of software systems. They serve as a vital mechanism for detecting bugs, improving code design, and facilitating knowledge sharing among team members. Despite their importance, code reviews can be challenging due to the programming expertise, domain knowledge, and most importantly, time required from the developers who are chosen to review the code. These reviews often impose a significant cognitive load on developers and are highly relying on the availability of senior developers, making them a bottleneck in the development life cycle. Moreover, the time spent reviewing code reduces the time senior developers can dedicate to their primary responsibilities, namely, developing new features and fixing bugs.

Generative AI and in particular Large Language Models have become very popular towards automating numerous tasks in software engineering ranging from writing of code to testing it. In this paper, we explore Large Language Models towards automatically generating code reviews for our code in Ericsson. In particular, we rely on a relatively light weight approach that avoided expensive pre-training and instruction fine tuning for generating code reviews. We experimented with various prompts to obtain a concise yet precise code review. We discovered that adding contextual information to the LLM's input prompt helped the LLM generate better reviews.

Our input consists of Java code lines that have been modified (to fix a bug or add a new feature). Our solution then utilizes static program analysis to extract the enclosing method of the modified lines, thus 
ensuring that proper context is passed to the LLM. Our aim is to provide developers timely and consistent feedback as they commit their code to version control systems such as Gerrit and Git. Thus, we integrate our automated code review into the workflow to ensure early detection of issues. 
Our solution aims to reduce the cognitive burden that code review imposes on experienced software developers and thus allowing them to focus on their typical primary responsibilities of writing code to add new features and/or fix bugs.


A simple but naive approach to automating the code review generation is to prompt the LLM with a text message such as ``Please generate a code review for the following code''.

In reality, we observed that such simple prompts do not work well in practice. Thus, generating meaningful reviews requires careful consideration of the following factors: 
\begin{enumerate}
\item[\textbullet] \textbf{What data to prompt on?} The data, the code diff in our case, must be passed to LLM with the right context. The context is defined as the code changeset and the reference method or the function on which the changes were made. The context in our solution is termed as the enclosing method. 
\item[\textbullet] \textbf{How to prompt?} We experimented with multiple prompting styles within the constraint window size of LLMs. The prompting styles ensure that the reviews are short, crisp, and human-like (typically LLMs generate verbose reviews); that the reviews are related to the enclosing method (LLMs hallucinate and generate random function names, objects, etc.); and, that no code is generated in the output (LLMs do this by default).
\item[\textbullet] \textbf{How to validate prompt results?} Engineering various prompting styles was not sufficient enough to generate good reviews. We further developed a validation mechanism to fine-tune the reviews generated, including summarizing and Ranking reviews, storing reviews generated by LLMs and validating them through human experts, and re-calibrating prompts using expert feedback.
\end{enumerate}

We also considered the following practical aspects when developing our approach:
\begin{itemize}
    \item It generates relevant, concise, and accurate reviews, ensuring consistency with code changes while minimizing mistakes. 
    \item It is fast, cost-efficient, and easy to integrate with existing development workflows while maintaining security, logging feedback for continuous improvement, and incorporating human validation where necessary.
\end{itemize}

This paper presents the approach and preliminary results of an evaluation (survey-based), offering insights into its practical application and potential improvements. We show that our approach can significantly enhance the quality of code reviews, helping to alleviate the cognitive load on human reviewers and streamlining the development process. 

The remainder of the paper is organized as follows: Section \ref{sec:related}  outlines literature related to our work. Section \ref{sec:design} describes the design used in our research. Section \ref{sec:prompts} lists the prompts we used. Section \ref{sec:solution} outlines our approach. Section \ref{sec:evaluation} shows the preliminary results of evaluating our approach. Section \ref{sec:discussion} synthesizes the analyze of our preliminary results. Finally, in Section \ref{sec:conclusion}, we conclude the paper and discuss a research roadmap for our work in this area.

\section{Related Work}
\label{sec:related}
Automated code review has been a key focus in software engineering, aiming to improve software quality while reducing the manual effort of reviewing code. Traditional approaches rely on manual reviews and static analysis tools such as SonarQube and Fortify, which are often time-consuming, prone to error and limited in scope to certain types of fixes and improvements \cite{cihan2024automated}. 

With the advent of Large Language Models (LLMs), researchers have explored code review methods based on them. Studies have shown the feasibility of using LLMs for code review tasks. Recent work has also highlighted the benefits of LLM-based tools in detecting bugs, identifying code smells, and optimizing software development workflows \cite{rasheed2024ai}. Another study \cite{yu2024github}  explored the use of Github Copilot and ChatGPT for AI suggestions, but found that real-world adoption remains limited due to concerns regarding accuracy, computational cost, and scalability.  In the paper by Fang et al. \cite{fang2024llm}, models are trained on large code repositories and review data, enabling them to provide context-aware suggestions that improve code quality beyond static analysis tools.

A study \cite{yu2024fine} explores how fine-tuned LLMs can generate more readable and understandable review comments, and studies the potential of LLMs to assist human reviewers by providing relevant explanations for detected issues. A literature review \cite{zubair2024repair} examines the use of LLMs in program repair, which overlaps with code review by identifying and suggesting fixes for bugs. They analyzed 41 studies, using mostly open-access datasets, and found that encoder-decoder architectures are good for such tasks. Another review on LLMs for code generation tasks \cite{chen2024survey} discusses the need for repository-level evaluation, mentioning continuous integration testing and code review simulations as methods to assess LLM performance in real-world scenarios. This suggests that LLMs could integrate with version control systems for comprehensive code analysis.
A study \cite{pornprasit2024fine} explored fine-tuning models on specialized code review datasets, and found this approach required significant computational resources and was not necessarily cost effective or feasible for organizations with limited budgets. They also compared fine-tuning with zero-shot and few-shot prompting and found that few-shot learning without personas offers a balance between performance and efficiency. 
A study evaluating the impact of an LLM-based automated review tool in industry settings \cite{cihan2024automated} found that while LLM-assisted reviews improved bug detection, they also led to increased pull request closure times due to redundant or inaccurate suggestions. Moreover, the effectiveness of LLM-generated reviews is significantly dependent on model selection, prompt engineering, and integration with existing development workflows \cite{weber2024productivity}.

In contrast to \cite{pornprasit2024fine}, ours is a lightweight approach not relying on expensive fine tuning which requires annotated data, computation and so on. Compared to \cite{cihan2024automated}, where they explored a commercial black-box tool for code review within a software development organization, we explore a lightweight home-grown approach to code review in practice at Ericsson.


\section{Research Design}
\label{sec:design}
Our research employs a structured methodology to develop and evaluate an automated code review system using large language models (LLMs). The study follows a three-step approach, comprising problem identification, solution development, and evaluation.

\begin{figure}[htbp]
\centerline{\includegraphics[width=\columnwidth]{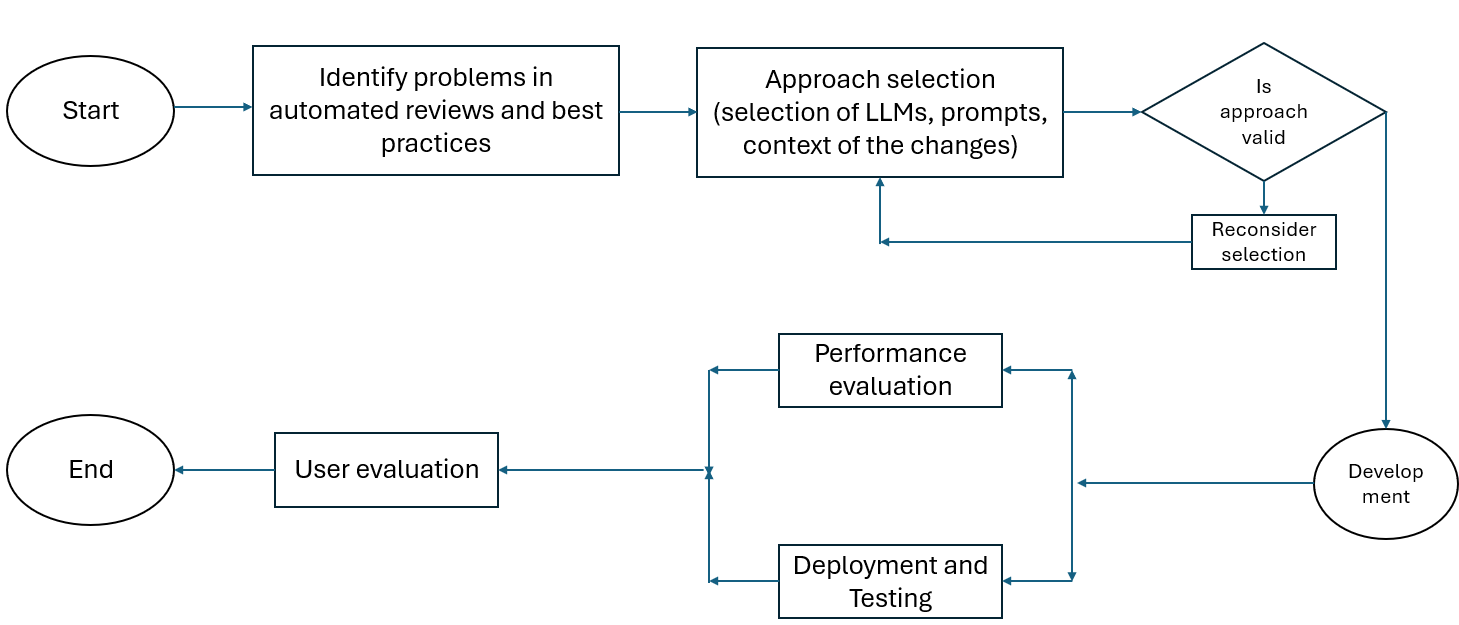}}
\caption{Research design structure.}
\label{fig:researchdesign}
\end{figure}

\subsection{Problem Identification and Scope Definition}
To begin, we analyzed the challenges faced by developers in performing manual code reviews, focusing on determining a method by which code reviews can be generated at scale and cheaply, given the limited availability of expert developers, as well as the limitations of existing static analysis tools. We conducted a literature review by searching for the following keywords in academic search engines like google scholar: "automated code review using LLMs", "review code using large language models", "LLM in software engineering" and "extracting the bounding functions of diff lines in a change request". Our objective was to identify gaps in existing solutions for automated code review, both in academic and industrial contexts. Based on the results, we formulated our research questions, focusing on how LLMs can enhance the quality and efficiency of code reviews while addressing constraints such as cost and latency.

\subsection{Solution Development}
We designed and implemented an automated code review pipeline that integrates open-source LLMs, such as Llama, mixtral and Code Llama, with program analysis techniques. Our approach consisted of contextual understanding of code changes by extracting the focal method or enclosing function for the diff lines involved in each change before sending to the LLM, providing relevant LLM prompts, and incorporating a mechanism to refine the generated reviews. The system is integrated with widely used version control platforms such as Gerrit, as well as tools such as VS Code by writing custom plugins, to provide seamless review capabilities.

\subsection{Evaluation Methodology}
To assess the effectiveness of our solution, we conducted a survey-based evaluation with experienced developers from different teams, having at least 3 years of experience and at the position of senior software engineer or above. The study involved presenting LLM-generated reviews for code snippets from actual Ericsson projects and gathering expert feedback on relevance, accuracy, and usability. 
For the pairwise comparison task, the experts were given two LLM generated code reviews for the same code snippet: consisting smaller and bigger LLM models. They were asked to compare the two code reviews as to which is better. The experts were not informed which reviews were from the smaller models. 
We also collated the feedback from users using the code review tool, who used the buttons provided along with the tool to give their feedback on the quality of the generated code reviews. 


The structure of our research design is illustrated in figure ~\ref{fig:researchdesign}. 

\section{Code Review Prompts}
\label{sec:prompts}
In this section, we describe the actual prompts we used for the code review to the LLM. 
\begin{enumerate}
\item[\textbullet] Simple Code Review Prompt: Provide a succinct analysis of the code snippet below. Only offer comments if significant concerns are identified, ensuring brevity without vagueness. Do not describe the functionality of the code. Avoid generating new code. Focus solely on critical evaluation. If the code is satisfactory, refrain from commenting.
\item[\textbullet] Detailed Code Review Prompt: As a code reviewer, conduct a thorough analysis of the provided code snippet to identify any significant issues, including but not limited to:
runtime errors and edge cases, logic flaws and potential bugs, algorithm correctness, gaps in error handling, architecture and design patterns, naming conventions and readability, performance concerns, maintainability issues. If any critical issues are discovered, regardless of category, provide a concise review in approximately 200 words. If no issues are found, please state this explicitly.

\item[\textbullet] Security Code Review Prompt: A version that focuses on security related issues. 

\item[\textbullet] Few-Shot Prompt: Create prompt templates for few-shot code reviews for Java and Python, tailored to the modified lines and examples provided. 

\item[\textbullet] Issue Topics Prompt: In the role of a reviewer for a recent pull request within a large-scale real-time telecommunication system, evaluate the code snippet below for potential concerns related to:Code Design: Issues with algorithms, data operations, function calls, object creation, and operation sequencing.Code I/O: Concerns related to input/output or GUI. Code Logic. 
Additionally, if quality-related issues beyond these concerns are present, indicate their existence in the code.
\end{enumerate}

\begin{figure}[htbp]
\centerline{\includegraphics[width=60pt]{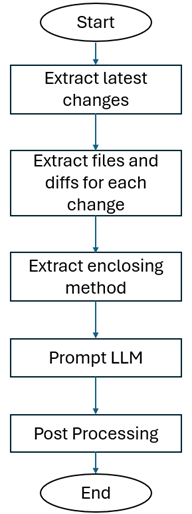}}
\caption{Steps of the code review pipeline.}
\label{fig:pipeline}
\end{figure}

\section{Solution Overview}
\label{sec:solution}

The code review pipeline is illustrated in figure ~\ref{fig:pipeline}. The steps of the end to end code review pipeline are as follows: 
\begin{enumerate}
\item Extract latest changes: We first get a list of the latest changes from the Gerrit API and show in our user interface (UI). Developers can choose a specific changeset in the UI, after which the system will fetch all the changes and revisions for the chosen changeset. 
\item Extract files and diffs for each change: We call the Gerrit API to extract a list of files for each change. We then generate and save the files. Finally, we call Gerrit APIs to get the diff lines for each change and reconstruct the line numbers. 
\item Extract enclosing method: Our code then \emph{extracts} the enclosing Java function for each diff, given the diff lines with the line numbers. We use the popular Tree-Sitter parser for Java to parse the code, build the abstract syntax tree, visit each method declaration and thus find the method(s) that enclose the modified lines.~\footnote{https://tree-sitter.github.io/tree-sitter/}
\item Prompt the LLM: Once we have the enclosing function, we feed it to the LLM such as code llama with a suitable prompt to perform a code review for the function. 
\item Post-processing: After we get the code review output from the LLM, we perform post-processing steps such as presenting, saving and summarizing the reviews. 
\end{enumerate}

\begin{figure}[htbp]
\centerline{\includegraphics[width=\columnwidth]{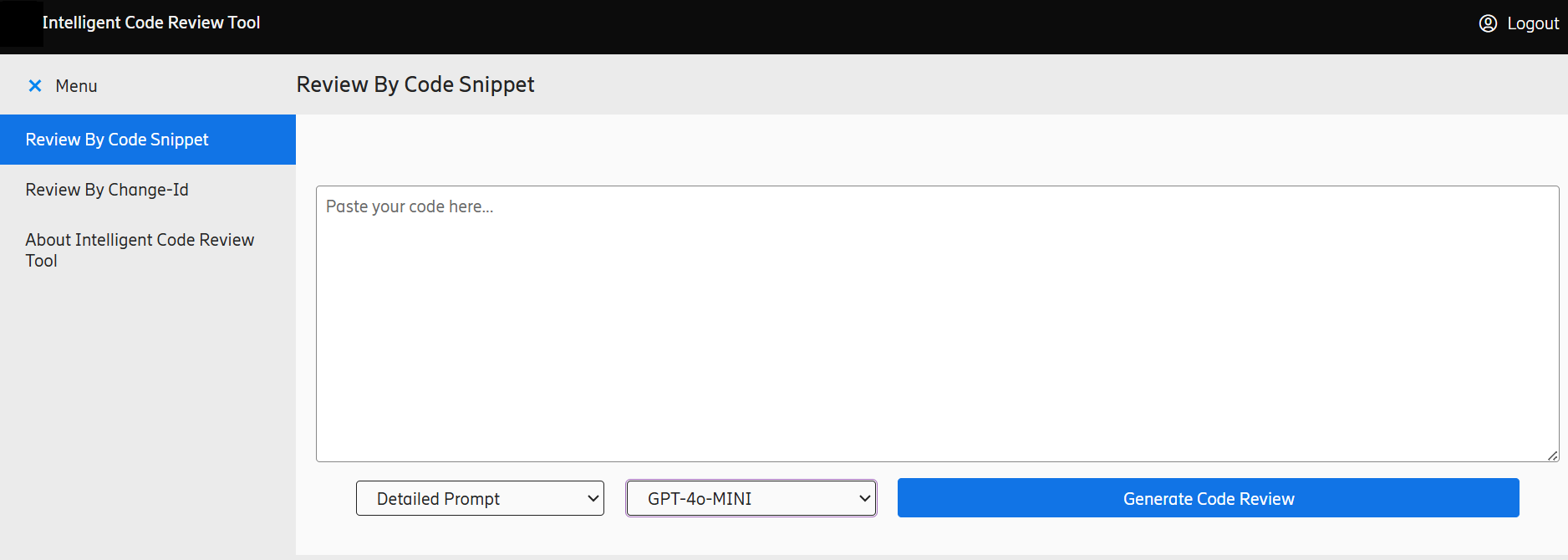}}
\caption{Web based user interface for the code review tool.}
\label{fig:crtui}
\end{figure}

A screenshot of the web-based user interface of the code review tool is shown in figure~\ref{fig:crtui}. The user interface allows developers to seamlessly incorporate automated code reviews into their workflow.

\begin{figure}[htbp]
\centerline{\includegraphics[width=0.8\columnwidth]{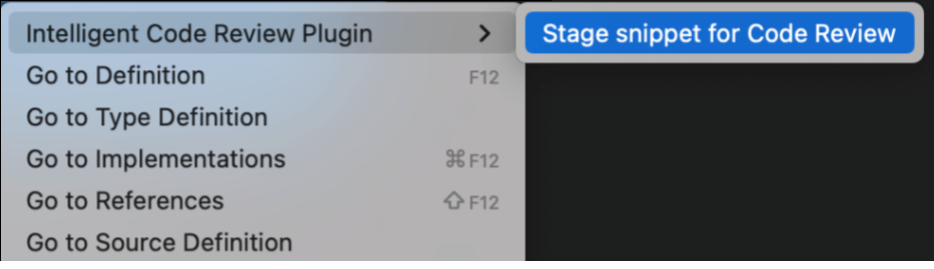}}
\caption{Screenshot of the popup menu in the VS Code plugin of the code review tool.}
\label{fig:plugin}
\end{figure}

The code review tool has been incorporated as a plugin in Visual Studio Code. A screenshot of the pop-up menu of the plugin is shown in figure~\ref{fig:plugin}. 

\section{Evaluation}
\label{sec:evaluation}
We have evaluated the effectiveness of our solution for automated code review with the help of user surveys with expert developers within our organization. Our three research questions were to analyze how good the LLMs are in generating reviews, the understand which LLMs produced better reviews and how good the code review tool was in practice. 

\subsection{RQ1: How good is the code review generation by the LLMs?}
Our survey design was as follows: 
We had a form with 10 code snippets from Ericsson code repository. In the form, each code snippet was displayed along with the full code path and LLM generated reviews for that code. The experts were invited to provide their comments for each of these 10 reviews: such as which parts of the review are relevant and which parts are not. The members of our team were available throughout the survey to address any questions. Finally, the experts' responses were compiled for further analysis. 
9 expert developers were invited for the survey, out of which 8 completed the first round. We included 10 Java methods, varying in length between long, medium and short methods. The methods were shared alongside the corresponding LLM reviews. The survey was designed to be finished in less than an hour. Free text reviews were also collected. A static prompt was employed for the code snippets, asking the LLM to "write a critical code review for following method, include only things to improve in less than 50 words, and do not generate code".

Qualitative reviews: We obtained a mix of positive, negative and neutral comments from the experts. The number of each type of comments is shown in Table \ref{tab:survey1}. 

\begin{table}[ht]
\centering
\caption{Results of a survey of experts on the quality of code reviews}
\begin{tabular}{lccc}
\hline
Code Snippets & Positive & Neutral & Negative \\ 
\hline
Long Snippets   & 3      & 2          & 4 \\ 
Medium Snippets & 2      & 2          & 3 \\ 
Short Snippets  & 1      & 4          & 3 \\
\hline
\end{tabular}
\label{tab:survey1}
\end{table}

A sampling of the reviews is as follows: 
\begin{enumerate}
\item[\textbullet] Positive comments: Experts appreciated suggestions like creating meaningful variable names and noted that reviews generally met expectations.
\item[\textbullet] Neutral comments: Feedback was mixed.  Some experts felt that method explanations were unnecessary or misplaced, and noted that important static fields were not addressed.
\item[\textbullet] Negative comments: The experts criticized instances where the LLM generated wrong variable types and input parameters, where the reviews were not relevant or completely incorrect, had missing abstractions, or were too verbose. 
\end{enumerate}

\begin{table}[ht]
\centering
\caption{Results of the pairwise comparison task in the survey of experts}
\begin{tabular}{lccc}
\hline
\textbf{LLM A versus LLM B} & \textbf{Votes for A} & \textbf{Votes for B} \\
\hline
Code LLaMA 13B vs LLaMA 2 13B         & 2 & 1 \\
Code LLaMA 13B vs Code LLaMA 7B       & 2 & 1 \\
Code LLaMA 13B vs LLaMA 2 7B          & 3 & 2 \\
LLaMA 2 13B vs Code LLaMA 7B          & 4 & 1 \\
LLaMA 2 13B vs LLaMA 2 7B             & 2 & 1 \\
Code LLaMA 7B vs LLaMA 2 7B           & 1 & 2 \\
\hline
\end{tabular}
\label{tab:pairwise}
\end{table}

\subsection{RQ2: Which LLM produced the best review?}
In this pairwise comparison task, experts were given reviews from 13B and 7B smaller LLM models, such as llama 2 13B and code llama 7B. The task involved 9 experts, each making 4 comparisons, totaling 36 pairwise evaluations across four models: Llama 2 13B, Code Llama 13B, Llama 2 7B, and Code Llama 7B. Table~\ref{tab:pairwise} shows the results of the pairwise comparison task. The initial results of the experts indicate that the Code llama 13B model seemed to be better than the others. 

\subsection{RQ3: How good is the code review tool in practice?}
We conducted a third survey of nine expert developers who were requested to use our code review tool in their day to day activities for fifteen days. The participants were then asked to answer a number of questions about how useful they found the tool. The questions and their answers are depicted below.
\begin{enumerate}
\item[\textbullet]\textbf{Did the code review tool save you time?}
Four of the nine developers agreed that the tool saved them time.
\item[\textbullet]
\textbf{Did the code review tool improve your overall coding efficiency?}
Four people agreed that it was effective or highly effective, two said it was somewhat effective, while three felt it was not very effective. 
\item[\textbullet]
\textbf{When the review generated by the tool is not satisfactory, what is lacking in it?}
Four said that the review was sometimes just explaining the code, two said the reviews were factually incorrect and three said that the review was focusing on irrelevant areas. 
\item[\textbullet]
\textbf{How frequently have you been using the code review tool?}
Two said that they were using it on a regular basis, while five said sometimes and the remaining two said they used the tool only once. 
\end{enumerate}

To conclude, our user surveys with experts are encouraging. It has provided valuable insights into the strengths and weaknesses of our tool for automated code review, guiding further refinement and development of the tool.

We also conducted a few other experiments to evaluate the effectiveness of the LLM in generating code reviews. First was about adversarial prompting, where we introduced a few bugs in the code snippets and fed to the LLM. For 10 short and 5 medium code snippets, the LLM was able to find bugs in all the cases, but these were relatively easy to catch logical bugs. Then, for 5 change IDs, we checked if the LLM left comments on unchanged lines in the bounding function, and found that it sticks to comments only about the changed lines. Finally, we measured the time it took for the LLM to generate reviews on large (100-300 lines) code snippets, and it was around 6 seconds. Interestingly, for short code snippets also the LLM took a similar time, around 5-6 seconds. 

\section{Discussion}
\label{sec:discussion}
Our study presents a simple approach to automated code review using LLMs. We extract the focal methods for the lines changed in change requests,
which is a simple and fast way to provide the context to the
LLM, and use inexpensive open source LLMs (like llama 2) tailored for code review. Our solution can easily integrate into existing systems for code development. Our approach ensures fast reviews without increasing API usage costs, making it practical for large-scale projects. Real-world developer surveys validate that our method reduces redundant feedback, and enhances usability compared to existing tools. 

For researchers, our study highlights the potential of bounding method-based LLM approaches in automated code review, opening avenues for self-refining AI reviews, prompt optimization, and domain-specific model adaptation. For industry practitioners, our findings demonstrate that LLM-powered code review can be scalable, cost-efficient, and easily integrated into existing workflows without extensive model fine-tuning, helping teams improve review speed, reduce manual effort, and enhance code quality.



\section{Conclusion and Research Roadmap}
\label{sec:conclusion}
In this paper, we have presented an LLM based system that utilizes static program analysis for automated code reviews. We include the modified lines (diff) along with the enclosing method obtained by program analysis and feed to the LLM with a prompt to generate code reviews.
We evaluated our system on our Ericsson code base by using expert developers as the oracle.
We believe that our initial results are promising.

Our work does not require the expensive fine tuning that certain state of the art works require nor does it rely on an external black box tool which we cannot use due to security concerns.

This work is an ongoing research effort aimed at improving accuracy, efficiency, and real-world adoption of LLM-based automated code reviews. In the next phases of our research, each of about six months duration, we have a research roadmap aimed to explore new advancements and make short term improvements.

We describe in some detail our future plans below:
\begin{enumerate}
\item[\textbullet]In the first phase, we plan to create a framework for ongoing user surveys with a broader sample of developers and expand prompt experimentation with new and upcoming LLMs and prompting strategies such as zero-shot, few-shot, and chain-of-thought. 
\item[\textbullet]In the second phase we intend to explore Retrieval-Augmented Generation (RAG) and Graph-RAG by retrieving relevant documentation, past review decisions, and dependency graphs before generating feedback. 
\item[\textbullet]The third phase will focus on developing a multi-agent framework using a system such as Crew AI, where specialized agents operate autonomously in parallel and handle different review tasks, continuously learning from feedback and past review data to improve the review quality and enforce consistency. 
\item[\textbullet]The final phase will focus on integrating the system with various internal tools for software engineering in our organization. 
\end{enumerate}


\begin{thebibliography}{9}

\bibitem{chen2024survey}
L. Chen, Q. Guo, H. Jia, Z. Zeng, X. Wang, Y. Xu, J. Wu, Y. Wang, Q. Gao, J. Wang, and W. Ye, "A survey on evaluating large language models in code generation tasks," \textit{arXiv preprint arXiv:2408.16498}, Aug. 2024.

\bibitem{cihan2024automated}
U. Cihan, V. Haratian, A. Ic¸oz, M. K. G ¨ ul, D. ¨ O. Devran, E. F. Bayendur, B. M. Uc¸ar, and E. Tuz¨ un, "Automated Code Review In Practice," \textit{arXiv preprint arXiv:2412.18531}, Dec. 2024.

\bibitem{fang2024llm}
C. Fang, N. Miao, S. Srivastav, J. Liu, R. Zhang, R. Fang, R. Tsang, N. Nazari, H. Wang, and H. Homayoun, "Large language models for code analysis: Do LLMs really do their job?" in \textit{33rd USENIX Security Symposium (USENIX Security 24)}, pp. 829-846, 2024.

\bibitem{pornprasit2024fine}
C. Pornprasit and C. Tantithamthavorn, "Fine-tuning and prompt engineering for large language models-based code review automation," \textit{Information and Software Technology}, vol. 175, p. 107523, Nov. 2024.

\bibitem{rasheed2024ai}
Z. Rasheed, M. A. Sami, M. Waseem, K. K. Kemell, X. Wang, A. Nguyen, K. Systa¨, and P. Abrahamsson, "Ai-powered code review with llms: Early results," \textit{arXiv preprint arXiv:2404.18496}, Apr. 2024.

\bibitem{weber2024productivity}
T. Weber, M. Brandmaier, A. Schmidt, and S. Mayer, "Significant productivity gains through programming with large language models," \textit{Proceedings of the ACM on Human-Computer Interaction}, vol. 8, no. EICS, pp. 1-29, Jun. 2024.

\bibitem{yu2024github}
X. Yu, L. Liu, X. Hu, J. W. Keung, J. Liu, and X. Xia, "Where Are Large Language Models for Code Generation on GitHub?," \textit{arXiv preprint arXiv:2406.19544}, Jun. 2024.

\bibitem{yu2024fine}
Y. Yu, G. Rong, H. Shen, H. Zhang, D. Shao, M. Wang, Z. Wei, Y. Xu, and J. Wang, "Fine-tuning large language models to improve accuracy and comprehensibility of automated code review," \textit{ACM Transactions on Software Engineering and Methodology}, vol. 34, no. 1, pp. 1-26, Dec. 2024.

\bibitem{zubair2024repair}
F. Zubair, M. Al-Hitmi, and C. Catal, "The use of large language models for program repair," \textit{Computer Standards and Interfaces}, vol. 103951, Nov. 2024.



\end{thebibliography}
\end{document}